\documentclass[prb,twocolumn,showpacs,preprintnumbers,amsmath,amssymb]{revtex4}

\usepackage{graphicx}
\usepackage{bbm}
\usepackage{float}

\begin{document}

\preprint{}
\title{Ferroelectricity in the Dion-Jacobson CsBiNb$_2$O$_7$ from first principles}
\author{Craig J. Fennie and Karin M. Rabe}
\affiliation{Department of Physics and Astronomy, Rutgers University,
        Piscataway, NJ 08854-8019}
\date{\today}

\begin{abstract}
We have studied ferroelectricity in Dion-Jacobson 
CsBiNb$_2$O$_7$ from first principles. Using group-theoretical 
analysis and first-principles density functional calculations of 
the total energy and phonons, we perform a systematic study of the 
energy surface around a paraelectric prototypic phase. Our results suggest
that CsBiNb$_2$O$_7$ is a ferroelectric with a polarization of 
P$_s$=40$\mu$C cm$^{-2}$. We propose further experiments to clarify 
this point.

\end{abstract}


\maketitle


The ABO$_3$ perovskites with their corner-shared octahedra, display an
amazing variety of physical properties by simply changing the A and B 
cations. In the rational design of new materials much effort has been 
directed towards manipulating the connectivity of perovskites by forming
structures consisting of 2-d layers of perovskites separated by cation 
network. Homologous series of these layered perovskites such as the 
Dion-Jacobson A[A'$_{n-1}$B$_n$O$_{3n+1}]$, Ruddlesden-Popper 
A$_2$[A'$_{n-1}$B$_n$O$_{3n+1}$], and Aurivillius 
[Bi$_2$O$_2$] [ A'$_{n-1}$B$_n$O$_{3n+1}$] (where A=alkali metal, A' 
alkaline earth, and B transition metal) have been formed by a number of 
novel growth techniques including molecular beam epitaxy~\cite{schlom.mse.01}
and soft chemistry techniques~\cite{schaak.chemmat.02} such as ion-exchange and 
intercalation reactions.

While ferroelectricity is ubiquitous in the ABO$_3$ perovskite titanates, 
niobates, and tantalates, the only layered perovskites to display a
ferroelectric (FE) transition has been the Aurivillius compounds,
while a predicted Ruddlesden-Popper FE~\cite{fennie.prb.05a} 
has yet to be confirmed experimentally. Recently, Lightfoot and coworkers 
performed neutron diffraction on ceramic samples of the n=2 Dion-Jacobson 
compound CsBiNb$_2$O$_7$ and found that the room temperature structure 
crystallizes in the polar space group P2$_1$am.~\cite{snedden.jssc.03} 
Nevertheless, from electrical measurements of the static dielectric constant,
they concluded that CsBiNb$_2$O$_7$ does not display ferroelectricity. 

In this letter we perform group theoretical analysis and first-principles 
calculations of the structural energetics and phonons of CsBiNb$_2$O$_7$. 
We show that the polar structure observed in Ref.~\onlinecite{snedden.jssc.03}
is a stable low-symmetry phase of CsBiNb$_2$O$_7$ for which we calculate
a substantial polarization, P$_s$$\approx$40$\mu$C cm$^{-2}$. Further we 
point out that the observed polar distortions from a non-polar reference 
structure are consistent with the criteria established by Abrahams~\cite{abrahams.acta.88}
for systems having a high probability of displaying a FE transition. All
this strongly suggests that CsBiNb$_2$O$_7$ is a FE.

\begin{figure}[b]
\includegraphics[scale=0.25]{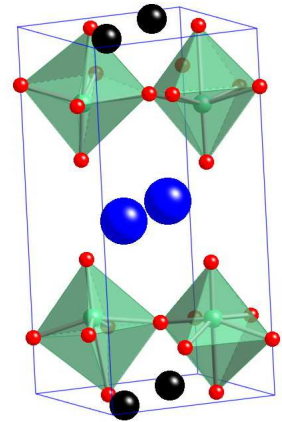}\\
\caption{\label{fig:groups} Structure of ferroelectric CsBiNb$_2$O$_7$.}
\end{figure}

First-principles density-functional calculations using projector 
augmented-wave potentials were performed within the local density 
approximation (LDA) as implemented in the {\it Vienna ab initio 
Simulation Package} ({\sf VASP}).~\cite{VASP,PAW} The wavefunctions
were expanded in plane waves up to a kinetic energy cutoff of 500 eV. 
Integrals over the Brillouin zone were approximated by sums on a $6 \times 6 \times 2$ 
$\Gamma$-centered $k$-point mesh. Polarization was calculated using the modern theory 
of polarization~\cite{king-smith.prb.93} as implemented in {\sf VASP}.

We approach the problem of identifying stable, low-symmetry phases of 
CsBiNb$_2$O$_7$ and possible ferroelectricity by investigating the 
lattice instabilities of a hypothetical reference structure.~\cite{hypothetical,rabe.04} 
Reference structures of this type can be chosen by detecting a 
pseudosymmetry~\cite{igartua.prb.96} of the observed polar phase, 
that is by finding structures in a group-subgroup relation where 
the atomic distortions relating the two phases are relatively small. 
The simple tetragonal (paraelectric) space group P4/mmm, which the 
related compound, CsLaNb$_2$O$_7$ has been shown to form at room 
temperature~\cite{kumada.acta.96} satisfies this requirement. While it is 
reasonable to assume that at a high temperature CsBiNb$_2$O$_7$ transforms to 
P4/mmm,  we must emphasize that regardless of whether this transition occurs, 
P4/mmm is a good reference structure.

 From x-ray and neutron powder diffraction, Ref.~\onlinecite{snedden.jssc.03} 
showed that the unit cell of CsBiNb$_2$O$_7$ was orthorhombic at room 
temperature and in a $\sqrt{2}$$a_t$ $\times$ $\sqrt{2}$$a_t$ $\times$
$c_t$ relation to the simple tetragonal reference structure, systematic 
reflections being consistent with the non-polar space group Pmam and with 
the polar space group P2$_1$am.  Subsequent Rietveld refinements gave a 
better fit to the P2$_1$am space group. 

\begin{figure}[t]
\includegraphics[scale=0.3]{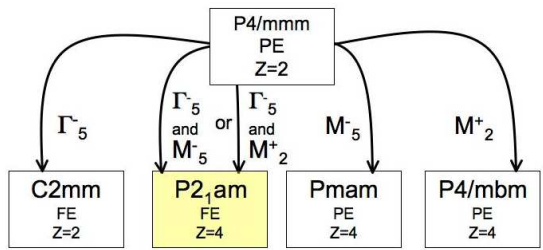}\\
\caption{\label{fig:group} Unstable phonons of the prototypic phase, P4/mmm, connecting 
possible low-temperature paraelectric (PE) and ferroelectric (FE) phases.}
\end{figure}

Both Pmam and P2$_1$am are subgroups of P4/mmm. Group-theoretic analysis
shows that the zone-center $\Gamma$-phonons
and the zone-boundary $M$-point phonons drive all relevant 
transitions.~\cite{bilbao,isotropy} Performing phonon calculations at 
these selected points in the Brillouin zone and subsequently investigating
the structural energetics around the reference structure facilitates a
systematic method of determining possible low-symmetry structures. 
Specifically, as seen from Fig.~\ref{fig:group}, Pmam is related to P4/mmm by the 
freezing-in of a single M$^-_5$ zone-boundary phonon. On the other hand, 
a $\Gamma^-_5$ zone-center ferroelectric mode and a M-point zone-boundary 
mode are required to account for the symmetry change associated with P2$_1$am.

We performed full optimization of the lattice parameters and internal coordinates 
of CsBiNb$_2$O$_7$ in the tetragonal reference structure space group P4/mmm. Our 
calculated structural parameters are displayed in Table~\ref{table:structure} where
they are written in terms of the $\sqrt{2}$$a_t$ x $\sqrt{2}$$a_t$ x $c_t$ lattice 
vectors so as to readily compare with the P2$_1$am structure. Next we calculated 
the $\Gamma^-_5$ phonons by computing the relevant block of the dynamical matrix 
from finite differences of the Hellmann-Feynman forces. We found one unstable 
$\Gamma^-_5$ phonon with frequency $\omega$= i 185 cm$^{-1}$. The character 
of the real-space eigendisplacement of this unstable mode consists mostly of Bi-ions
moving against oxygen in Bi-O planes as would be expected due to the Bi 6s lone
pair~\cite{seshadri} but also consists of a smaller amount of Nb-O displacements.
The freezing-in of this ferroelectric mode 
would lead directly to the C2mm space group without a doubling of the unit cell. 
Next we computed the M$^-_5$, M$^+_2$, and M$^+_3$ phonons at the q=(1/2, 1/2, 0) 
point of the Brillouin zone. M$^-_5$ has two unstable modes with frequencies  
$\omega_1$= i 210 cm$^{-1}$ and $\omega_2$= i 70 cm$^{-1}$. The M$^-_5$ 
modes are antiferrodistortive, the character of which are  Bi-O and Nb-O
displacements for $\omega_1$ and $\omega_2$ respectively. The freezing-in of these 
M$^-_5$ modes leads to the paraelectric, doubled unit cell, Pmam space group. Finally,
the M$^+_2$ and the M$^+_3$ phonons both consist only of oxygen motion in the Nb-O planes.
The M$^+_2$ phonon is unstable ($\omega$= i 190 cm$^{-1}$) leading to space 
group P4/mbm while the M$^+_3$ phonon is stable.

\begin{table}[t]
\caption{Crystal structure of ferroelectric CsBiNb$_2$O$_7$,
Space Group: $P$2$_1$$am$, $a$= 5.412\,\AA\/,\,$b$=5.343\,\AA\/,\,$c$= 11.22\,\AA\/ }
\begin{ruledtabular}
\begin{tabular}{lccc}

 & &P4/mmm&P2$_1$am\\ \hline

\begin{tabular}{l} Atom\\Cs\,(2b)\\Bi\,(2a)\\Nb\,(4c)\\O1\,(2a)\\O2\,(4c)
\\O3\,(4c)\\O4\,(4c)\\ \end{tabular}
&

&\begin{tabular}{ccccc} x&&y&&z\\$1\over4$&&$1\over4$&&$1\over2$\\$1\over4$&&$1\over4$&&0
\\$1\over4$&&$3\over4$&&0.2077\\ $1\over4$&&$3\over4$&&0\\$1\over4$&&$3\over4$&&0.3617
\\0&&0&&0.1631\\$1\over2$&&$1\over2$&&0.1631\\
\end{tabular}

&\begin{tabular}{ccccc} x&&y&&z\\0.2536&&0.2609&&$1\over2$\\0.2999&&0.2652&&0\\0.2414&&0.7525&&0.2077\\
0.2065&&0.6741&&0\\0.2496&&0.7842&&0.3642\\0.0174&&0.0383&&0.1551\\ 0.4190&&0.4380&&0.1888\\
\end{tabular}

\end{tabular}
\end{ruledtabular}
\label{table:structure}
\end{table}

Based on the results of the phonon calculations,  we see that transitions out 
of the P4/mmm reference structure to the space groups C2mm, Pmam, P4/mbm, and
P2$_1$am are all possible due to the freezing-in of one or more unstable modes,
as summarized in Fig.~\ref{fig:group}.
To identify which phase is most energetically favorable and therefore most likely
to be observed at low temperatures, we performed a series of structural relaxations
within each of the above mentioned space groups. We relaxed all lattice parameters
and internal coordinates, the results are as follows:
\begin{eqnarray}\nonumber
\begin{tabular}{ccccr}
Space Group	&&Irrep & &$\Delta$ E/ f.u. \\ 
P4/mbm	        && M$^+_2$        & &-400 meV \\
C2mm	        && $\Gamma^-_5$   &&-550 {meV} \\
Pmam	        && M$^+_5$        &&-350 {meV} \\
P2$_1$am	&&$\Gamma^-_5$ and M$^+_2$ & & \\
&or& $\Gamma^-_5$ and M$^-_5$ &&-800 {meV} \\
\end{tabular}
\end{eqnarray}
where it is clear that P2$_1$am, being 800 meV lower than the tetragonal 
reference structure and 450 meV lower than Pmam, is the computed ground 
state, in full agreement with the Rietveld refinements of 
Ref.~\onlinecite{snedden.jssc.03}.
The results of the structural relaxations within the P2$_1$am space group 
are shown in Table~\ref{table:structure}. We find lattice constants 
$\approx$1$\%$ smaller than those determined experimentally at 300 K, typical of LDA
calculations. In Table~\ref{table:bonds} we compare with the experimentally 
determined bond lengths. We find excellent agreement, within 1-2$\%$ except 
for the Bi-O$_3$ bonds which differ by 3-5$\%$.  

\begin{table}[t]
\caption{Bond lengths (\AA) of CsBiNb$_2$O$_7$,
Space Group: $P$2$_1$$am$, Exp.: $a$= 5.495\AA,\,$b$=5.423\AA,\,$c$= 11.38\AA.
LDA: $a$= 5.412\AA,\,$b$=5.343\AA,\,$c$= 11.22\AA }
\begin{ruledtabular}
\begin{tabular}{lcc}

Atoms& Exp. Ref.~\onlinecite{snedden.jssc.03}\footnotemark[1] &LDA\\ \hline

Cs\,\,-\,\,O$_{(2)}$& 3.038&2.968\\
                    & 3.144&3.096\\
                    & 3.202&3.133\\
                    & 3.232&3.184\\
\\
Nb\,\,-\,\,O$_{(1)}$& 2.403&2.375\\
Nb\,\,-\,\,O$_{(2)}$& 1.755&1.764\\
Nb\,\,-\,\,O$_{(3)}$& 1.941&1.957\\
                    & 2.024&2.037\\ 
Nb\,\,-\,\,O$_{(4)}$& 1.973&1.948\\
                    & 2.082&2.031\\
\\
Bi\,\,-\,\,O$_{(1)}$& 2.257&2.224\\
                    & 2.269&2.243\\
                    & 3.243&3.199\\
                    & 3.256&3.228\\
Bi\,\,-\,\,O$_{(3)}$& 2.707&2.614\\
                    & 2.794&2.654\\
Bi\,\,-\,\,O$_{(4)}$& 2.409&2.398\\
                    & 3.311&3.354\\
\end{tabular}
\end{ruledtabular}
\footnotetext[1]{Powder neutron diffraction, room temp.}
\label{table:bonds}
\end{table}

From phonon calculations, we have shown that lattice instabilities of a 
high symmetry reference structure of CsBiNb$_2$O$_7$ lead naturally to a 
P2$_1$am ground state. As we have stated the symmetry of this space group 
allows a ferroelectric polarization to develop along the $a$-axis. In fact, 
from Berry-phase calculations of the polarization we find that CsBiNb$_2$O$_7$ 
should have a polarization P$_s$$\approx$ 40 $\mu$C cm$^{-2}$, which is about 
three times larger than the polarization of the $n$=2 Aurivillius compound 
SrBi$_2$Ta$_2$O$_9$ and comparable to that of the $n$=3 Bi$_4$Ti$_3$O$_{12}$.

In principle, a material is considered a ferroelectric if both 1) it displays a 
spontaneous polarization (P$_s$) and 2) if P$_s$ can be flipped in an electric
field. This definition is reflected in a experimentally derived structural 
criteria for identifying possible ferroelectrics as 1) the largest 
$\Delta$x $>$ 0.1 \AA\ and 2) all $\Delta$x $<$ 1.0 \AA, where $\Delta$x can be
though of as an atomic distortion along the polar axis from the centrosymmetric 
reference structure.~\cite{abrahams.acta.88} 
In CsBiNb$_2$O$_7$, the largest atomic distortion relative to the oxygen framework
is $\Delta$x(Bi)$\approx$0.5\AA\ which clearly meets Abrahams' structural criteria
and suggests, independent of the first-principles results, that CsBiNb$_2$O$_7$
is a FE.

Experimentally, an indication of a ferroelectric transition is usually shown by
the strong temperature dependence of the static dielectric constant around T$_c$,
normally a Curie-Weiss behavior. At temperatures much lower than T$_c$, a strong
temperature dependence of the dielectric constant is not expected, for example, 
the static dielectric constant of LiNbO$_3$ (T$_c$= 1450 K) over the temperature 
range 0K-300K. The fact that the static dielectric constant of CsBiNb$_2$O$_7$ 
does not display a large temperature variation around room temperature does not 
preclude a ferroelectric transition at a higher temperature. Considering
the PE-to-FE energy scale that we calculated, $\Delta$E$\approx$-400meV/f.u.,
T$_c$ is expected to be much 
higher than room temperature. This is consistent with other known ferroelectrics
where the stereochemical activity of the Bi lone pair is the driving force 
towards ferroelectricity such as in BiFeO$_3$~\cite{fennie.bifeO3} and in 
SrBi$_2$Ta$_2$O$_9$.~\cite{perez.prb.04} In fact, from an empirical 
expression that has proven quite successful in predicting T$_c$ of several new 
ferroelectrics, we find $T_c=2\times 10^4 (\Delta x)^2 \,\,K=810\,\,K$   
where $\Delta$x is the largest metal atom (in this case Bi) displacement along 
the ferroelectric direction after a shift of origin so that 
$\Delta$x(Cs)+$\Delta$x(Bi)+$\Delta$x(Nb)=0.~\cite{abrahams.acta.01}
We suggest that direct measurements of the polarization in CsBiNb$_2$O$_7$ be performed, 
such as E-P hysteresis loops, to reveal it's ferroelectric nature since the actual
T$_c$ may in fact be higher than the decomposition temperature of the material.

We have confirmed that the ground state of CsBiNb$_2$O$_7$ forms 
in the polar space group P2$_1$am and expect a substantial polarization 
P$_s$ $\approx$ 40 $\mu$C cm$^{-2}$. The similarity of the computed lattice 
instabilities to the $n$=2 Aurivillius compounds~\cite{perez.prb.04} is
intriguing, the implications of which should be pursed further given the 
fundamental and technological interest in such materials. The Dion-Jacobson
family of compounds may provide another avenue towards the design of new 
ferroelectrics and, given the propensity towards ion-exchange, 
multiferroics.~\cite{shen.06}

Useful discussions with Ram Seshadri and  D.H$.$ Vanderbilt 
are acknowledged. This work was supported by NSF-NIRT Grant No. DMR-0103354.
CJF acknowledges the support of Bell Labs and Rutgers University.



\begin{thebibliography}{10}
\bibitem{hypothetical} 
By hypothetical reference we mean a CsBiNb$_2$O$_7$ structure that has yet 
to be observed experimentally.
\end{thebibliography}
\end{document}